# The INTERSPEECH 2020 Deep Noise Suppression Challenge: Datasets, Subjective Testing Framework, and Challenge Results

*Chandan K. A. Reddy*[1], *Vishak Gopal*[1], *Ross Cutler*[1], *Ebrahim Beyrami*[1], *Roger Cheng*[1], *Harishchandra Dubey*[1], *Sergiy Matusevych*[1], *Robert Aichner*[1], *Ashkan Aazami*[1], *Sebastian Braun*[2], *Puneet Rana*[1], *Sriram Srinivasan*[1], *Johannes Gehrke*[1]

[1]Microsoft Corporation, Redmond, WA, USA

{Chandan.karadagur, firstname.lastname}@microsoft.com

## Abstract

The INTERSPEECH 2020 Deep Noise Suppression (DNS) Challenge is intended to promote collaborative research in real-time single-channel Speech Enhancement aimed to maximize the subjective (perceptual) quality of the enhanced speech. A typical approach to evaluate the noise suppression methods is to use objective metrics on the test set obtained by splitting the original dataset. While the performance is good on the synthetic test set, often the model performance degrades significantly on real recordings. Also, most of the conventional objective metrics do not correlate well with subjective tests and lab subjective tests are not scalable for a large test set. In this challenge, we open-sourced a large clean speech and noise corpus for training the noise suppression models and a representative test set to real-world scenarios consisting of both synthetic and real recordings. We also open-sourced an online subjective test framework based on ITU-T P.808 for researchers to reliably test their developments. We evaluated the results using P.808 on a blind test set. The results and the key learnings from the challenge are discussed. The datasets and scripts can be found here for quick access https://github.com/microsoft/DNS-Challenge.

## 1. Introduction

As the number of people working remotely and in open office environments continues to increase, the desire to have a video/audio call with excellent speech quality and intelligibility has become more important than ever before. The degradation of speech quality due to background noise is one of the major sources for poor quality ratings in voice calls. The conventional Speech Enhancement (SE) techniques are based on statistical models estimated from the noisy observations. These methods perform well on stationary noises but fail to effectively suppress non-stationary noises [1]–[5]. Recently, SE is treated as a supervised learning problem in which the patterns within speech and noise are learned using the training data [6]. Deep Neural Networks (DNN) are used to estimate speech in either the spectral or time domain. DNN based methods are shown to outperform conventional SE techniques in suppressing non-stationary noises [7]–[10].

Most of the published literature reports experimental results based on objective speech quality metrics such as Perceptual Evaluation of Speech Quality (PESQ) [11], Perceptual Objective Listening Quality Analysis (POLQA) [12], Virtual Speech Quality Objective Listener (ViSQOL) [13], Speech to Distortion Ratio (SDR). These metrics are shown to not correlate well with subjective tests [14]. Few papers report subjective lab test results, and many that do are either not statistically significant, or the test set is small.

For a SE task, the training set is composed of noisy and clean speech pairs. Noisy speech is usually synthesized by mixing clean speech and noise. Testing the developed models on the synthetic test set gives a heuristic on model performance, but it is not enough to ensure good performance when deployed in real-world conditions. The developed models should be tested on representative real recordings of noisy speech from diverse noisy and reverberant conditions in which speech and noise are captured at the same microphone in similar acoustic conditions. It is hard to simulate these conditions using synthetic data as clean speech and noise signals are captured independently in different acoustic environments. This makes it difficult for researchers to compare published SE methods and pick the best ones as there is no common test set that is extensive and representative of real-world noisy conditions. Also, there is no reliable subjective test framework that everyone in the research community could use. In [14], we open sourced the Microsoft Scalable Noisy Speech Dataset (MS-SNSD)[1] and an ITU-T P.800 subjective evaluation framework [14]. MS-SNSD includes clean speech and noise recordings and scripts to synthesize noisy speech with augmentation for generating the training set. In addition, a disjoint test set is provided for evaluation. But the test set was missing real recordings and not enough noisy conditions with reverberation. In addition, the P.800 implementation in [14] is missing some crowdsourcing features in P.808 such as hearing and environmental tests, trapping questions, and validation.

The Deep Noise Suppression (DNS) challenge is designed to unify the research work in SE domain by open sourcing the train/test datasets and subjective evaluation framework. We provide a large clean speech and noise datasets that are 30 times bigger than MS-SNSD [14]. These datasets are accompanied with configurable scripts to synthesize the training sets. Participants could use any datasets or augment their datasets of their choice for training. A part of the test set was released for the researchers to use during development. The other half was used as a blind test set to decide the final competition winners. We also open sourced the implementation of an online subjective evaluation framework using ITU-T P.808 [15]. We also provide the model and inference script for a recent SE method as a baseline algorithm for comparison [16].

---

[1] https://github.com/microsoft/MS-SNSD

In this paper, we will describe the datasets, subjective evaluation framework, and the baseline method. Finally, we will discuss the datasets used by the participants and the results from the DNS Challenge.

## 2. Datasets

The goal of releasing the clean speech and noise datasets is to provide researchers with the extensive and representative datasets to train their SE models. Previously, we released MS-SNSD [14] with a focus on extensibility. In recent years, the amount of audio data available over the internet has exploded due to increased content creation on YouTube, smart devices, and audiobooks. Though most of these datasets are useful for tasks such as training audio event detectors, automatic speech recognition (ASR) systems, etc., most of the SE models need a clean reference, which is not always available. Hence, we synthesize noisy-clean speech pairs.

### 2.1. Clean Speech

The clean speech dataset is derived from the public audiobooks dataset called Librivox[1]. Librivox corpus is available under the permissive creative commons 4.0 license [17]. Librivox has recordings of volunteers reading over 10,000 public domain audiobooks in various languages, with majority of which are in English. In total, there are 11,350 speakers. Many of these recordings are of excellent speech quality, meaning that the speech was recorded using good quality microphones in a silent and less reverberant environments. But there are many recordings that are of poor speech quality as well as with speech distortion, background noise, and reverberation. Hence, it is important to clean the data set based on speech quality.

We used the online subjective test framework ITU-T P.808 [15] to sort the book chapters by subjective quality. The audio chapters in Librivox are of variable length ranging from few seconds to several minutes. We randomly sampled 10 audio segments from each book chapter, each of 10 seconds in duration. For each clip, we had 2 ratings, and the MOS across all clips was used as the book chapter MOS. Figure 1 shows the results, which show the quality spanned from very poor to excellent quality.

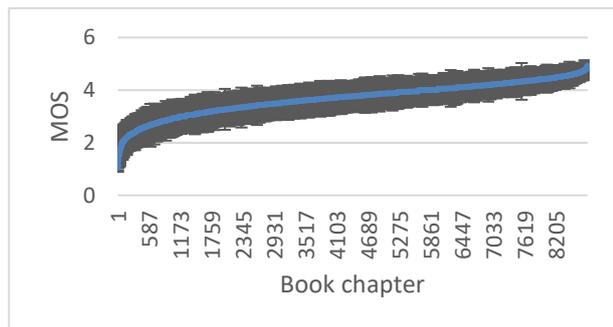

Figure 1: Sorted Librivox P.808 MOS quality with 95% confidence intervals

The upper quartile with respect to MOS was chosen as our clean speech dataset, which are the top 25% of clips with MOS as a metric. The upper quartile is comprised of audio chapters with $4.3 \leq MOS \leq 5$. We removed clips from speakers with less than 15 minutes of speech. The resulting dataset has 500 hours of speech from 2150 speakers. All the filtered clips are then split into segments of 10 seconds. In total, we use approx. 441 hours of clean speech for generating the upper quartile subset.

### 2.2. Noise Dataset

The noise clips were selected from Audioset[2] [18] and Freesound[3]. Audioset is a collection of about 2 million human-labeled 10s sound clips drawn from YouTube videos and belong to about 600 audio events. Like the Librivox data, certain audio event classes are overrepresented. For example, there are over a million clips with audio classes music and speech and less than 200 clips for classes such as toothbrush, creak, etc. Approximately 42% of the clips have a single class, but the rest may have 2 to 15 labels. Hence, we developed a sampling approach to balance the dataset in such a way that each class has at least 500 clips. We also used a speech activity detector to remove the clips with any kind of speech activity. This is to avoid training the noise suppression model to learn to suppress speech. The resulting dataset has about 150 audio classes and 60,000 clips. We also augmented an additional 10,000 noise clips downloaded from Freesound and DEMAND databases [19]. The chosen noise types are more relevant to VoIP applications.

### 2.3. Noisy Speech

The clean speech and noise datasets can be found in the repo[4]. The noisy speech database is created by adding clean speech and noise at various Signal to Noise Ratio (SNR) levels. We compute segmental SNR using segments in which both speech and noise are active. This is to avoid overshooting of amplitude levels in impulsive noise types such as door shutting, clatter, dog barking, etc. We synthesize 30s long clips by augmenting clean speech utterances and noise. The SNR levels are sampled from a uniform distribution between 0 and 40 dB. The mixed signal is then set to target Root Mean Square (RMS) level sampled from a uniform distribution between -15 dBFS and -35 dBFS. The data generation scripts are open sourced in the DNS-Challenge repo[5].

### 2.4. Test clips

#### 2.4.1. Development stage test set

We open sourced a new test set comprising both synthetic and real recordings. It is a general practice to evaluate the SE method on a synthetic test set. But a synthetic test set is not representative of what we observe in the wild. The synthetic test set might be useful in tuning the model during the development phase using objective metrics such as PESQ and POLQA that require a clean reference. Generally, in synthetic data, the original clean speech and noise are collected in different acoustic conditions using two different microphones and are mixed to form noisy speech. With real recordings, the clean

---

[1] https://librivox.org/
[2] https://research.google.com/audioset/
[3] https://freesound.org/
[4] https://github.com/microsoft/DNS-Challenge/tree/master/datasets
[5] https://github.com/microsoft/DNS-Challenge

speech and noise are captured at the same microphone and acoustic conditions.

The test set is divided into 4 categories with 300 clips in each:
1. Synthetic clips without reverb
2. Synthetic clips with reverb
3. Real recordings collected internally at Microsoft
4. Real recordings from AudioSet

For synthetic test clips, we used Graz University's clean speech dataset [20] which consists of 4,270 recorded sentences spoken by 20 speakers. For the synthetic clips with reverb, we add reverberation to the clean files using the room impulse responses recorded internally at Microsoft with RT60 ranging from 300ms to 1300ms. We sampled 15 clips from 12 noise categories we deem highly important for VoIP scenarios to synthesize 180 noisy clips. The 12 categories are fan, air conditioner, typing, door shutting, clatter noise, car, munching, creaking chair, breathing, copy machine, baby crying and barking. The remaining 120 noise clips were randomly sampled from the remaining 100+ noise classes. The SNR levels were sampled from a uniform distribution between 0 dB and 25 dB. The real recordings collected internally at Microsoft consist of recorded noisy speech in various open office and conference rooms noisy conditions. We hand-picked 300 audio clips with speech mixed in noise from AudioSet that we felt are relevant to audio calls we experience in noisy conditions.

*2.4.2. Blind test set*

The blind test set also comprised of 600 clips (300 synthetic and 300 real recordings). This set was used for the final evaluation. The synthetic test clips are generated in a similar way as that of the development stage test set, except for using unseen clean speech, noise, and room impulse responses.

We crowdsourced the real recordings data collection process using Amazon Mechanical Turk (MTurk). The MTurk participants captured their speech in a variety of noisy acoustic conditions. They also used a variety of devices (headphones and speakerphones) to record their clips. This gave us a test set with a diverse noisy speech in realistic conditions.

## 3. Baseline SE method

As a baseline, we will use the recently developed SE method from [16] which is based on Recurrent Neural Network (RNN). For ease of reference, we will call this method as Noise Suppression Net (NSNet). This method uses log power spectra as input to predict the enhancement gain per frame using a learning machine based on Gated Recurrent Units (GRU) and fully connected layers. Please refer to the paper for more details of the method.

NSNet is computationally efficient. It only takes 0.16ms to enhance a 20ms frame on an Intel quad-core i5 machine using the ONNX run time v1.1[1]. It is subjectively evaluated using a large test set showing improvement over a conventional SE method. We have open sourced the inference script and the model in ONNX format in the challenge DNS-Challenge repo[2].

## 4. Online Subjective Evaluation Framework ITU-T P.808

We use the ITU-T P.808 Subjective Evaluation of Speech Quality with a Crowdsourcing Approach [15] methodology to evaluate and compare SE methods using Absolute Category Ratings (ACR) to estimate a Mean Opinion Score (MOS). We created an open source[3] implementation of P.808 using the MTurk platform. This system has the following features/attributes:

- Raters are first qualified using a hearing and environmental test before they can start rating clips. This ensures raters have a sufficient hearing ability, a good quality listening device, and a quiet environment to do ratings in. Our implementation allows raters to start rating clips immediately after being qualified which increased the rating speed by ~5X compared to having a separate qualification stage.
- Raters are given several training examples but are not screened using the results; the training is used for anchoring purposes.
- Audio clips are rated in groups of clips (e.g., N=10). Each group includes a gold clip with known ground truth (e.g. a clean or very poor clip) and a trapping question (e.g., "This is an interruption: Please select option 2"). The gold and trapping questions are used for filtering out "spam" raters who are not paying attention.
- Every hour raters are also given a comparison rating test using gold samples (e.g., which is better, A or B) to verify their environment is still valid to do ratings in.
- Raters are restricted to rating a limited number of clips per P.808 recommendations to reduce rater fatigue.

To validate the measurement system accuracy, we rated the ITU Supplement 23 Experiment 3 [21] dataset which has published lab-based MOS results. The system gives a Spearman correlation coefficient of 0.90 to the lab results given in ITU Supplement 23 (MOS is computed per test condition). To validate the system repeatability, we ran the ITU Supplement 23 twice (on separate days, with <10% overlapped raters, and $1/10^{th}$ the ratings as Run 1) and the results were similar (see [21]).

Table 1: P.808 Spearman rank correlation with ITU Supplement 23 Experiment 3

|  | $\rho$ |
| --- | --- |
| ITU Supplement 23 Run 1 | 0.93 |
| ITU Supplement 23 Run 2 | 0.87 |

## 5. DNS Challenge Tracks

Every participating SE method will fall in one of the two tracks depending on the computational complexity. Track 1 is focused on low computational complexity for Real-Time applications. The algorithm must take less than $T/2$ (in ms) to process a frame of size $T$ (in ms) on an Intel Core i5 quad-core machine clocked at 2.4 GHz or equivalent processors. The frame length $T$ must be less than or equal to 40ms. Track 2 is Non-Real-Time track and does not have any constraints on computational time so that researchers can explore deeper models to attain exceptional speech quality. In both the tracks, the SE method can have a maximum of 40ms look ahead. To infer the current frame $i$ (in ms), the algorithm can access any number of past frames but only 40ms of future frames ($i$+40ms).

---

[1] https://github.com/microsoft/onnxruntime
[2] https://github.com/microsoft/DNS-Challenge/tree/master/NSNet-baseline
[3] https://github.com/microsoft/P.808

Three winners were selected from each track based on the subjective speech quality evaluated on the blind test set using ITU-T P.808 framework.

## 6. Datasets used and Challenge Results

### 6.1. Datasets used

The DNS challenge allowed participants to use any dataset of their choice to train their models. Participants found the DNS Challenge dataset useful and used the datasets as it was provided. A few teams added more clean speech and noise data from other corpuses and augmented with the DNS Challenge data. Many teams added reverberation to clean speech using the image method and attempted to do simultaneous dereverberation and noise suppression. A few teams modified the configurable scripts provided in our repo to meet their needs. Many teams chose discrete SNR and target levels instead of randomly sampling from a range of values. The total number of hours used to train the models varied depending on their models and the availability of computational resources.

### 6.2. Challenge Results

We received 28 submissions from 19 teams. 9 teams participated in both the tracks. We conducted the subjective evaluations in two phases. In phase 1, we included all the 28 submissions plus noisy test set in one P.808 run with 10 raters per clip. This resulted in a 95% confidence interval (CI) of 0.02 across each submission. The results are shown in Table 2. The 'Team #' is the number assigned to the participants. Complexity indicates Real-Time (RT) or Non-Real-Time (NRT). 'Synthetic' corresponds to synthetic clips with no reverberation, 'Synthetic reverb' corresponds to synthetic clips with reverberation and the other column corresponds to real recordings. MOS value is the average of ratings per clip across each condition. 'dMOS' is the difference between the MOS after enhancement and MOS of the noisy blind set before enhancement. The wide span of dMOS values shows that we received a variety of models from participants.

Table 2: Phase 1 P.808 Results

| Team # | Complexity | Synth dMOS | Real dMOS | Reverb dMOS | Overall dMOS | 95% CI |
|---|---|---|---|---|---|---|
| 9 | NRT | 0.74 | 0.55 | 0.55 | 0.60 | 0.02 |
| 9 | RT | 0.59 | 0.53 | 0.38 | 0.51 | 0.02 |
| 29 | RT | 0.69 | 0.51 | 0.32 | 0.51 | 0.02 |
| 29 | NRT | 0.66 | 0.43 | 0.37 | 0.47 | 0.02 |
| 17 | NRT | 0.52 | 0.41 | 0.46 | 0.45 | 0.02 |
| 14 | NRT | 0.53 | 0.44 | 0.39 | 0.45 | 0.02 |
| 17 | RT | 0.54 | 0.42 | 0.43 | 0.45 | 0.02 |
| 30 | NRT | 0.48 | 0.36 | 0.24 | 0.36 | 0.02 |
| 20 | NRT | 0.43 | 0.31 | 0.34 | 0.35 | 0.02 |
| 37 | RT | 0.35 | 0.33 | 0.24 | 0.31 | 0.02 |
| 15 | NRT | 0.41 | 0.33 | 0.16 | 0.31 | 0.02 |
| 6 | NRT | 0.36 | 0.34 | 0.12 | 0.29 | 0.02 |
| 15 | RT | 0.31 | 0.27 | 0.16 | 0.25 | 0.02 |
| 18 | RT | 0.36 | 0.36 | -0.12 | 0.24 | 0.02 |
| 20 | RT | 0.27 | 0.22 | 0.20 | 0.23 | 0.02 |
| 22 | RT | 0.25 | 0.24 | 0.17 | 0.23 | 0.02 |
| 25 | NRT | 0.42 | 0.27 | -0.16 | 0.20 | 0.02 |
| 5 | NRT | 0.30 | 0.21 | 0.06 | 0.19 | 0.02 |
| 41 | NRT | 0.34 | 0.21 | 0.00 | 0.19 | 0.02 |
| 40 | RT | 0.21 | 0.20 | 0.14 | 0.19 | 0.02 |
| 5 | RT | 0.23 | 0.20 | 0.09 | 0.18 | 0.02 |
| 30 | RT | 0.17 | 0.13 | 0.12 | 0.14 | 0.02 |
| 41 | RT | 0.28 | 0.10 | -0.07 | 0.11 | 0.02 |
| 36 | RT | 0.14 | 0.07 | -0.07 | 0.05 | 0.02 |
| Baseline | RT | 0.17 | 0.03 | -0.14 | 0.02 | 0.02 |
| Noisy | | O (3.32) | O (2.97) | O (2.78) | O (3.01) | 0.02 |
| 26 | RT | -0.02 | -0.07 | -0.07 | -0.06 | 0.02 |
| 10 | RT | -0.21 | -0.37 | -0.67 | -0.41 | 0.02 |
| 33 | RT | -0.33 | -0.50 | -0.30 | -0.41 | 0.02 |
| 33 | NRT | -0.41 | -0.58 | -0.38 | -0.49 | 0.02 |

To rank the top 3 teams in each track, we conducted a phase 2 P.808 run with 10 raters on top 3 teams in the RT track and top 4 teams NRT track. We used Analysis of Variance (ANOVA) to pick the top teams based on statistical significance. For these top models, we combined the Phase 1 10 ratings with Phase 2 10 ratings. The total of 20 ratings per clip gives a 95% CI of 0.01 per model.

The results of top teams in both the tracks are shown in Table 3. Table 4 shows the p-Values between model pairs computed using ANOVA. We set 0.05 as the threshold to determine a statistically significant difference between the two models. P-Value less than 0.05 indicates statistically significant difference. All the teams in the RT track show statistically significant differences in speech quality. Hence, the three prizes can be easily picked based on speech quality. Teams 17 and 29 overlap for the 2nd place in NRT track. As per the DNS Challenge rules, if there are overlapping models, we use the computational complexity as a metric to pick the winner. The team with lower complexity will win the higher prize.

Table 3: Phase 2 P.808 Results

| Team # | Complexity | Synthetic MOS | Synthetic dMOS | Real Recordings MOS | Real Recordings dMOS | Synthetic Reverb MOS | Synthetic Reverb dMOS | Overall MOS | Overall dMOS | 95% CI |
|---|---|---|---|---|---|---|---|---|---|---|
| 9 | NRT | 4.07 | 0.94 | 3.40 | 0.57 | 3.19 | 0.54 | 3.52 | 0.67 | 0.01 |
| 29 | RT | 4.00 | 0.87 | 3.37 | 0.54 | 2.94 | 0.30 | 3.42 | 0.57 | 0.01 |
| 9 | RT | 3.87 | 0.74 | 3.38 | 0.55 | 2.97 | 0.32 | 3.39 | 0.54 | 0.01 |
| 17 | NRT | 3.83 | 0.70 | 3.28 | 0.45 | 3.15 | 0.51 | 3.38 | 0.53 | 0.01 |
| 29 | NRT | 3.90 | 0.77 | 3.34 | 0.52 | 2.96 | 0.31 | 3.38 | 0.53 | 0.01 |
| 17 | RT | 3.83 | 0.69 | 3.27 | 0.44 | 3.11 | 0.47 | 3.36 | 0.51 | 0.01 |
| 14 | NRT | 3.76 | 0.63 | 3.32 | 0.49 | 2.98 | 0.33 | 3.34 | 0.49 | 0.01 |
| Blind test set | | 3.13 | 0.00 | 2.83 | 0.00 | 2.64 | 0.00 | 2.85 | 0.00 | 0.01 |

Table 4: p-Values between models for Phase 2 P.808 results using ANOVA

| | 9 NRT | 17 NRT | 29 NRT | 9 RT | 17 RT | 29 RT |
|---|---|---|---|---|---|---|
| 9 NRT | 1.00 | | | | | |
| 17 NRT | 0.00 | 1.00 | | | | |
| 29 NRT | 0.00 | 0.27 | 1.00 | | | |
| 9 RT | 0.00 | 0.00 | 0.00 | 1.00 | | |
| 17 RT | 0.00 | 0.29 | 0.38 | 0.00 | 1.00 | |
| 29 RT | 0.00 | 0.01 | 0.02 | 0.00 | 0.00 | 1.00 |

| > 0.05 | Not Statistically Significant |
| < 0.05 | Statistically Significant |

## 7. Conclusion

The DNS challenge is designed to promote real-time single microphone noise suppression for exceptional subjective speech quality. The number of participants to the challenge exceeded our expectations. The participants found the open sourced DNS challenge datasets and P.808 subjective evaluation tool useful. We hope that the future developments in SE use the DNS challenge test set to evaluate models and report results on this common test set. This will immensely help readers to easily compare SE methods across publications.

In the future, we would investigate using P.835 to focus on the quality of speech and noise separately. We would like to create a speaker-specific (personalized) noise suppression challenge in the future. Finally, we will develop a no reference MOS predictor using DNS challenge results that can be used as an objective metric to quickly evaluate SE models.

## 8. Acknowledgements

The P.808 implementation was written by Babak Naderi.